\documentclass[useAMS,usenatbib,usegraphicx]{mn2e}

\usepackage{graphicx}

\title[New observable for gravitational lensed transits]
       {New observable for gravitational lensing effects during transits}
       
\author[S. Kasuya, M. Honda and R. Mishima]
{Shinta Kasuya\thanks{E-mail: kasuya@kanagawa-u.ac.jp}, 
Mitsuhiko Honda and Risa Mishima\\
Department of Information Sciences, Kanagawa University, Kanagawa 259-1293, Japan}

\begin{document}

\date{Accepted 2010 October 4. Received 2010 October 2; in original form 2010 September 10}

\pagerange{1863-1868}
\pubyear{2011}
\volume{411}

\maketitle

\label{firstpage}

\begin{abstract}
We investigate gravitational lensing effects of an extrasolar planet transiting its host star. 
We focus on the `rising spikes' of the light curve just before and after the transit, which is 
a peculiar feature of the gravitational lensing, and find that it could be a novel 
observable for determining physical parameters. Detectability of such an effect is also
discussed.
\end{abstract}


%%%%%%%%%%%%%%%%%%%%%%%%%%%%%%%%%%%%%%%%%%%
\section{Introduction}
Observations of decrease of starlight by a transit of an extrasolar planet is one of the
important method not only for finding the planet itself but for determining physical
parameters of the transiting planet \citep{Cassen2006}. Roughly speaking, there are two
observables in the transit light curve: Duration and depth. The latter gives the planet
radius if the star radius is known, while the former leads to the planet orbital radius 
if one neglects the size and mass of the planet for the transit duration and the
orbital velocity of the planet, respectively.

There is often observations of the orbital velocity and the period of the transit
using the Doppler effects in addition to the transit observation \citep{Cassen2006}. 
Hence the planet mass can be determined uniquely, 
since the inclination is known to be near edge-on from the transit

To date, hundreds of the extrasolar planets are observed by various methods mostly 
by Doppler method and transits\footnote{%%
See \texttt{http://exoplanet.eu/}.}, but those with a large orbital radius more 
than $\sim$10 au are limited by these methods. This is because one could see the 
transit practically only once, at most, in one's life time, and such planets are also difficult 
to detect with the Doppler method.

As the distance between the star and the planet becomes larger, 
the mass of the planet could in principle affect the transit light curve due to microlensing effect. 
Microlensing with occultation has been considered generally or in other context 
\citep{Maeder1973, Bromley1996, Marsh2001, Bozza2002, Agol2002, BT2002, Agol2003, SaGi2003, Lee2009}. 
It makes the depression shallower, and steepens the slope of the ingress and egress of the 
dip \citep{Marsh2001}. However, the former can be explained by a smaller planet radius, 
while the latter could be due to a faster transiting velocity. 

A peculiarity of the microlensing effect on
the transit light curve would be `rising spikes' just before and after the transit.\footnote{%%
This feature was mentioned for the case which occultation and microlensing
effects are equal in \citet{SaGi2003}, and investigated for large source case in \cite{Agol2003}.}%% 
This is what we focus on in the present article, and we show that the height of the spike
has different parameter dependence from other observables such as the depth of the
light curve. In particular, it could be measurable in high precision observations such 
as by {\it Kepler} or other future facilities.

The structure of the article is as follows. In the next section, we consider both microlensing 
and occultation simultaneously to obtain the magnification of the starlight. We see typical 
observables from the transit light curve, and define a novel observable as the `rising spike'
in the lensed case in Sec.~3, so that we consider three observables: The depth, the duration, 
and the rising spike. In Sec.4, we show that the section of isosurfaces of these three observables 
will determine the physical parameters of the planet: The mass, the radius, and the orbital radius. 
In Sec.~5, we include the limb darkening 
as an example effect which may obscure the rising spike. We conclude in Sec.~6. 
Appendices are devoted to the other sections of the isosurfaces of three observables, and
the other cases for lensing and occultation in the aligned situation.

%%%%%%%%%%%%%%%%%%%%%%%%%%%%%%%%%%%%%%%%%%%
\section{Microlensing and occultation}
The lens equation for a point mass is written as
\begin{equation}
\beta(\theta) = \theta - \frac{\theta_E^2}{\theta},
\label{lens-eq}
\end{equation}
where $\theta$ is an angular separation between the lens and the image of the source, while
$\beta$ is that between the lens and the source without lensing %(See Fig.~\ref{fig:lens_eq}). 
(See Fig.~1).
$\theta_E$ is the Einstein angle,
\begin{equation}
\theta_E = \sqrt{\frac{4GM_L}{c^2}\frac{D_{LS}}{D_{OS}D_{OL}}},
\label{E_radius}
\end{equation}
which represents an effective size of the lens. Here $D_{OS}$ and $D_{OL}$ are the
distance from the observer to the source and the lens, respectively. $D_{LS}$ is the
distance between the lens and the source. $M_L$ is the lens mass, $G$ is the gravitational 
constant, and $c$ is the speed of light. Solving Eq.(\ref{lens-eq}), we obtain the solutions
\begin{equation}
\theta_\pm = \frac{\beta\pm\sqrt{\beta^2+4\theta_E^2}}{2},
\end{equation}
where $\theta_{+(-)}$ is the angle of the image whose light comes from the far (near) side of
the lens, which we will call it $+ (-)$ path hereafter.

%%%%%%%%%%%%%%%%%%%%%%%%%%%%%%%
\begin{figure}
\label{fig:lens_eq}
\begin{center}
\includegraphics[width=0.7\columnwidth]{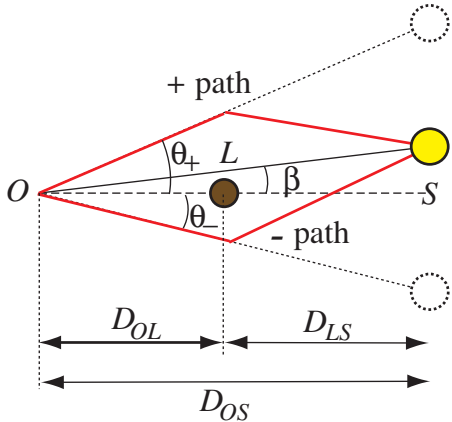}%{lens_eq.eps}
\caption{Schematic view of the geometry of the gravitational lensing.}
\end{center}
%\label{fig:lens_eq}
\end{figure}
%%%%%%%%%%%%%%%%%%%%%%%%%%%%%%%

The magnification of the light of a point on the source star through the $\pm$ path is given by
\begin{equation}
A_\pm = \frac{d\theta_\pm \theta_\pm d\phi}{d\beta \beta d\phi} 
= \frac{1}{2} \pm \frac{\beta^2+2\theta_E^2}{2\beta\sqrt{\beta^2+4\theta_E^2}},
\end{equation}
since the surface brightness is conserved. In the case of microlensing, one cannot
discriminate these two magnified images separately, so the observed magnification should
be the addition of the two as
\begin{equation}
A(\beta) = A_+ +|A_-| = \frac{\beta^2+2\theta_E^2}{\beta\sqrt{\beta^2+4\theta_E^2}}.
\end{equation}
Since the source is usually an extended object, integrating over the source, we can obtain 
the total magnification as
\begin{equation}
{\cal A} = \frac{\rm area \ with \ lens}{\rm area \ without \ lens}
=\frac{\int A(\beta) \beta d\beta d\phi}{\int \beta d\beta d\phi}.
\label{mag}
\end{equation}

So far we have considered only the point mass lens. Because we are interested in the
observation of the transit of a planet in front of the source star, we need to investigate the 
finite size effect of the lens planet. There are three cases: (1) Both $+$ and $-$ paths 
of the light from a certain point on the source star are occulted, (2) $-$ path is occulted, 
or (3) neither path is occulted. Therefore, when integrating over the source star in the numerator 
of Eq.(\ref{mag}), we must use $A(\beta)$ itself for case (3), $A_+(\beta)$ for case (2) instead 
of $A(\beta)$, or do not integrate over the region where case (1) holds.

In the simple configuration that the centres of the lens planet and the source star are 
aligned, one can calculate the magnification very easily.\footnote{%%
More general cases can be analytically estimated using elliptic integrals \citep{Agol2002}.}
The denominator of Eq.(\ref{mag}) is estimated as
\begin{equation}
\int_0^{2\pi} \int_0^{\theta_S} \beta d\beta d\phi = \pi \theta_S^2,
\label{denominator}
\end{equation}
as usual. Here $\theta_S = R_S/D_{OS}$, where $R_S$ is the radius of the 
source star. In order to calculate the numerator in the transit case when the occultation 
occurs, we must specify the condition to hold. The occultation takes place if $\beta(\theta_L)>0$, 
where $\theta_L=R_L/D_{OL}$ with $R_L$ being the radius of the lens planet 
(See case I in Fig.~\ref{fig:three}). It leads to
$\theta_L > \theta_E$, which implies that the size of the planet is bigger than that 
of the lens. Therefore, there is full occultation in the region $\beta< \beta(\theta_L)$,
while $-$ path is shielded for $\beta(\theta_L) < \beta < \theta_S$. Thus
the numerator of Eq.(\ref{mag}) is obtained as
\begin{eqnarray}
& & \hspace{-7mm}
\int_0^{2\pi} \int_{\beta(\theta_L)}^{\theta_S} A_+(\beta) \beta d\beta d\phi \nonumber \\
& & = \frac{\pi}{2}\left[\theta_S^2 - \beta_L^2 + \theta_S\sqrt{\theta_S^2+4\theta_E^2}
-\beta_L\sqrt{\beta_L^2+4\theta_E^2}\right],
\end{eqnarray}
where $\beta_L\equiv \beta(\theta_L)$. Therefore, we have
\begin{equation}
{\cal A} = \frac{1}{2}\left[ 1- \tilde{\beta}_L^2 + \sqrt{1+4\tilde{\theta}_E^2}
-\tilde{\beta}_L \sqrt{\tilde{\beta}_L^2+4\tilde{\theta}_E^2}\right], 
\label{amp}
\end{equation}
where the tilde denotes the variable normalised with respect to $\theta_S$ such that
$\tilde{\beta}_L=\beta_L/\theta_S$ and $\tilde{\theta}_E=\theta_E/\theta_S$. 
Notice that this magnification (or, more precisely, the diminution) corresponds to the depth at the
centre of the transiting light curve. For other situations, see Appendix A.

\begin{figure}
\includegraphics[width=0.9\columnwidth]{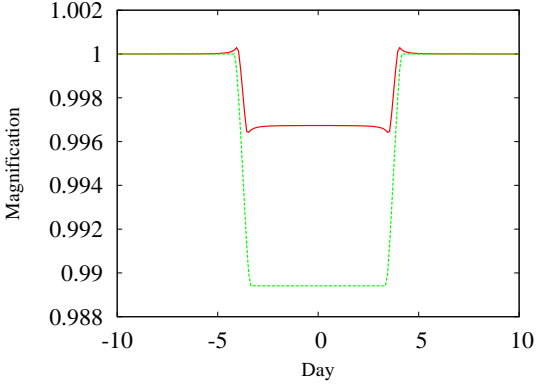}%{LC.eps}
\caption{Light curves of the transit. Red (solid) and green (dashed) lines represent the lensed
and unlensed cases, respectively. The vertical axis denotes the magnification ${\cal A}$, while
the horizontal axis shows the time in day from the centre of the transit in the case of the circular 
orbit. We take $M_L=10 M_J$, $R_L=R_J$, and $a=200$ au for the lens planet, and 
$M_S={\rm M}_\odot$, $R_S={\rm R}_\odot$, and $D_{OS}=10$~pc for the source star.}
\label{fig:LC}
\end{figure}

In order to follow the whole process of the transit, we integrate numerically the 
numerator of Eq.(\ref{mag}) to estimate the magnification. We show some example of the light 
curve in Fig.~\ref{fig:LC}. Here we take $M_L=10 M_J$, $R_L=R_J$, and $a=200$ au, where 
$M_J$ and $R_J$ are the Jupiter mass and radius, respectively. For parameters of the source
star, we set, here and hereafter, $M_S={\rm M}_\odot$, $R_S={\rm R}_\odot$, and 
$D_{OS}=10$~pc when calculating numerically. We should mention that 
$D_{OS}$-dependence is very weak for $D_{OS} \gg D_{LS}$. Also notice that, 
in this article, we restrict ourselves to the edge-on case 
in order to focus on the effect of lensing.

%%%%%%%%%%%%%%%%%%%%%%%%%%%%%%%
%%%%%%%%%%%%%%%%%%%%%%%%%%%%%%%

%%%%%%%%%%%%%%%%%%%%%%%%%%%%%%%%%%%%%%%%%%%
\section{Observables from the light curve and physical parameters}
\subsection{Unlensed transit} %%%%%%%%%%%

During the transit a light curve of a star drops down due to the occultation. We can
see how much it drops down and how long it lasts. These are the depth and duration
of the transit, respectively, and the only observables we can obtain from the light curve
(in a simplest case). If we measure the depth at the center of the transit, it is estimated as
\begin{equation}
\Delta =\frac{\theta_L^2}{\theta_S^2} = \frac{R_L^2}{R_S^2} \frac{D_{OS}^2}{D_{OL}^2}
=\frac{R_L^2}{R_S^2}\left( \frac{D_{OS}}{D_{OS}-a}\right)^2,
\label{delta_nolens}
\end{equation}
where $a\equiv D_{LS}$ is the orbital radius of the planet around the star. Once the 
parameters of the star ($R_L$ and $D_{OS}$) are known, it could be a function of 
$R_L$ and $a$: $\Delta=\Delta(R_L,a)$.
On the other hand, the duration of the transit is determined as
\begin{eqnarray}
& & \hspace{-7mm}
\tau = \frac{2(\theta_S+\theta_L)}{\omega_L} =\frac{2(\theta_S+\theta_L)}{v_L/D_{OL}} 
\nonumber \\
& & =2\left(R_L+\frac{D_{OS}-a}{D_{OS}}R_S\right)
\sqrt{\frac{a}{G(M_S+M_L)}},
\label{tau_nolens}
\end{eqnarray}
where the planet is assumed to revolve in a circular orbit around the mass center of the star 
and the planet, and has orbital velocity $v_L = \sqrt{G(M_S+M_L)/a}$. Again, if the 
parameters of the star ($M_L$, $R_L$, and $D_{OS}$) are known,
it becomes a function of $M_L$, $R_L$, and $a$: $\tau=\tau(M_L, R_L, a)$.
If $M_L$ could be much smaller than $M_S$ to be neglected, we can derived physical
parameters $R_L$ and $a$ from the transit observables $\Delta$ and $\tau$. 
Usually the transit observation is not the only information in our hands. Orbital velocity and/or
orbital period of the planet is obtained from the Doppler observation, and three physical
parameters are all determined.

\subsection{Lensed transit} %%%%%%%%%%%%%
Neither orbital velocity nor orbital period has been observed by Doppler method if the planet 
rotates around the star with relatively large orbital radius, typically farther than $\sim$10 au. 
In such a situation, the gravitational lensing effect becomes larger as the lens planet goes 
away from the star (See Eq.(\ref{E_radius})).

The gravitational lensing is known to affect the light curve of the transit in such a way that
the depth becomes shallower and the slopes of the ingress and the egress becomes
steeper, but the former effect on the light curve can be mimicked by the smaller planet radius,
while the latter may imply the faster orbital velocity of the planet. Therefore, one cannot
tell if it is really affected by the gravitational lensing. In order to confirm that one must surely
take into account the gravitational lensing effects, their peculiar feature should be observed
in the light curve. 

\begin{figure}
\includegraphics[width=0.9\columnwidth]{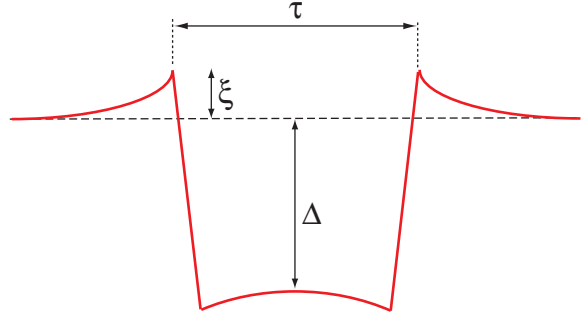}%{light_curve.eps}
\caption{Schematic view of the light curve of the lensed transit.}
\label{fig:LC_lens}
\end{figure}

Here we claim that the `rising spike' is the smoking gun of the gravitational lensing effect. 
It is the rise in the light curve just before the ingress (or just after the egress) of the transit,
shown as $\xi$ in Fig.~\ref{fig:LC_lens}. There are several key points that the rising spike
is a good observable for determining the physical parameters of the planet.
First of all, it is a unique consequence of the gravitational lensing effect, and cannot be 
mimicked by others. The effect could be observed in high precision observation such as
the {\it Kepler}, especially when the orbital radius is relatively large, where the orbital
velocity cannot be observed. For parameter determination, it has different parameter 
dependence from the depth or duration of the transit \citep{Agol2003}. 
Therefore, it is quite a unique observable.

%%%%%%%%%%%%%%%%%%%%%%%%%%%%%%%
%%%%%%%%%%%%%%%%%%%%%%%%%%%%%%%

\begin{figure}
\includegraphics[width=0.9\columnwidth]{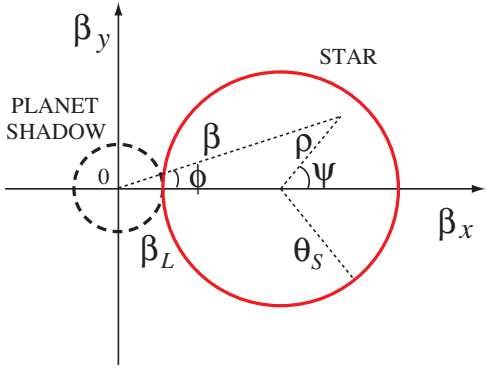}%{tangent.eps}
\caption{Geometry of the planet and the star at the moment of the peak of the rising spike.}
\label{fig:tangent}
\end{figure}

The amplitude of the rising spike can be calculated from the magnification of the light
at the moment when the lensed shadow of the planet has just been tangent to the star
as shown in Fig.~\ref{fig:tangent}. It is thus given by
\begin{equation}
1+\xi = \frac{1}{\pi} \int_0^{2\pi} \int_0^1A_+(\tilde{\beta}(\tilde{\rho},\psi)) 
\tilde{\rho} d\tilde{\rho} d\psi,
\end{equation}
where all the angular separations are normalized with respect to $\theta_S$ such as
$\tilde{\alpha} = \alpha / \theta_S$, and
\begin{equation}
A_+(\tilde{\beta}) = \frac{1}{2}\left[ 1+ \frac{\tilde{\beta}^2+2\tilde{\theta}_E^2}
{\tilde{\beta}\sqrt{\tilde{\beta}^2+4\tilde{\theta}_E^2}} \right],
\end{equation}
with
\begin{eqnarray}
\tilde{\beta} & = & \sqrt{\tilde{\beta}_x^2+\tilde{\beta}_y^2},\\
\tilde{\beta}_x & = & \tilde{\beta}_L + 1 + \tilde{\rho}\cos\psi, \\
\tilde{\beta}_y & = & \tilde{\rho}\sin\psi.
\end{eqnarray}
%%

%%%%%%%%%%%%%%%%%%%%%%%%%%%%%%%
%%%%%%%%%%%%%%%%%%%%%%%%%%%%%%%

Since the image of the source star is not occulted through $+$ path while fully occulted through
$-$ path, we can easily calculate $\xi$ using the magnification of the $+$ path for a point mass 
lens \citep{WM,Agol2002} as
\begin{eqnarray}
& & \hspace{-7mm}
\xi = \frac{1}{4\pi\sqrt{(\tilde{\beta}_L+2)^2(4\tilde{\theta}_E^2+\tilde{\beta}_L^2)}}
\nonumber \\
& & \hspace{3mm}
\times \left\{ (\tilde{\beta}_L+2)^2(4\tilde{\theta}_E^2+\tilde{\beta}_L^2) E(k) \right.
\nonumber \\
& & \hspace{6mm}
 -\left[ \tilde{\beta}_L^2(\tilde{\beta}_L+2)^2
 +8\tilde{\theta}_E^2(\tilde{\beta}_L^2+2\tilde{\beta}_L)\right] K(k) 
\nonumber \\
& & \hspace{6mm} \left.
+ 4\tilde{\beta}_L^2(\tilde{\theta}_E^2+1) \Pi(n,k) \right\} - \frac{1}{2},
\label{xi}
\end{eqnarray}
where $K(k)$, $E(k)$, and $\Pi(n,k)$ are the complete elliptical integrals of the 
first, second, and the third kinds, respectively, and 
\begin{equation}
n=1-\frac{\tilde{\beta}_L^2}{(\tilde{\beta}_L+2)^2}, \qquad
k=\sqrt{\frac{4\tilde{\theta}_E^2n}{4\tilde{\theta}_E^2+\tilde{\beta}_L^2}}.
\end{equation}
Since $\tilde{\theta}_E$ is a function of $M_L$ and $a$, and $\tilde{\beta}_L$ is a function of
$R_L$, $M_L$, and $a$, the rising spike $\xi$ is a function of 
$R_L$, $M_L$, and $a$: $\xi=\xi(R_L, M_L, a)$.

On the other hand, the depth and duration of the transit are modified due to the gravitational 
lensing effect. From Eq.(\ref{amp}), the depth is now estimated as
\begin{equation}
\Delta = 1 - {\cal A}
= \frac{1}{2}\left[1+\tilde{\beta}_L^2-\sqrt{1+4\tilde{\theta}_E^2}
+\tilde{\beta}_L\sqrt{\tilde{\beta}_L^2+4\tilde{\theta}_E^2}\right].
\label{delta}
\end{equation}
while the duration is given by
\begin{equation}
\tau = \frac{2(1+\tilde{\beta}_L)}{\tilde{\omega}_L}.
\label{tau}
\end{equation}
Therefore, these two observables are functions of $R_L$, $M_L$, and $a$: 
$\Delta=\Delta(M_L, R_L,a)$ and $\tau=\tau(M_L,R_L,a)$. Notice that they reduce to the 
unlensed case (\ref{delta_nolens}) and (\ref{tau_nolens}), respectively, for vanishing lens mass,
$M_L \rightarrow 0$, which implies $\theta_E \rightarrow 0$ and $\beta_L \rightarrow \theta_L$.

%%%%%%%%%%%%%%%%%%%%%%%%%%%%%%%%%%%%%%%%%%%
\section{Isosurfaces of the depth, the duration, and the rising spike}
Now we have the depth (\ref{delta}), the duration (\ref{tau}), and the rising spike (\ref{xi}),
which depend differently on the three physical parameters of the planet such as
the mass $M_L$, the radius $R_L$, and the orbital radius $a$. If these 
three observables are measured, we can reversely use these equations to determine
three physical parameters. Although it could be done in principle, the inverse problem is
practically difficult. However, we could numerically obtain the isosurfaces of $\Delta$, $\tau$,
and $\xi$ in three dimensions of $M_L$, $R_L$, and $a$. Intersection of these three
surfaces must be a point, the solution that we need.

\begin{figure}
\begin{tabular}{c}
\includegraphics[width=0.9\columnwidth]{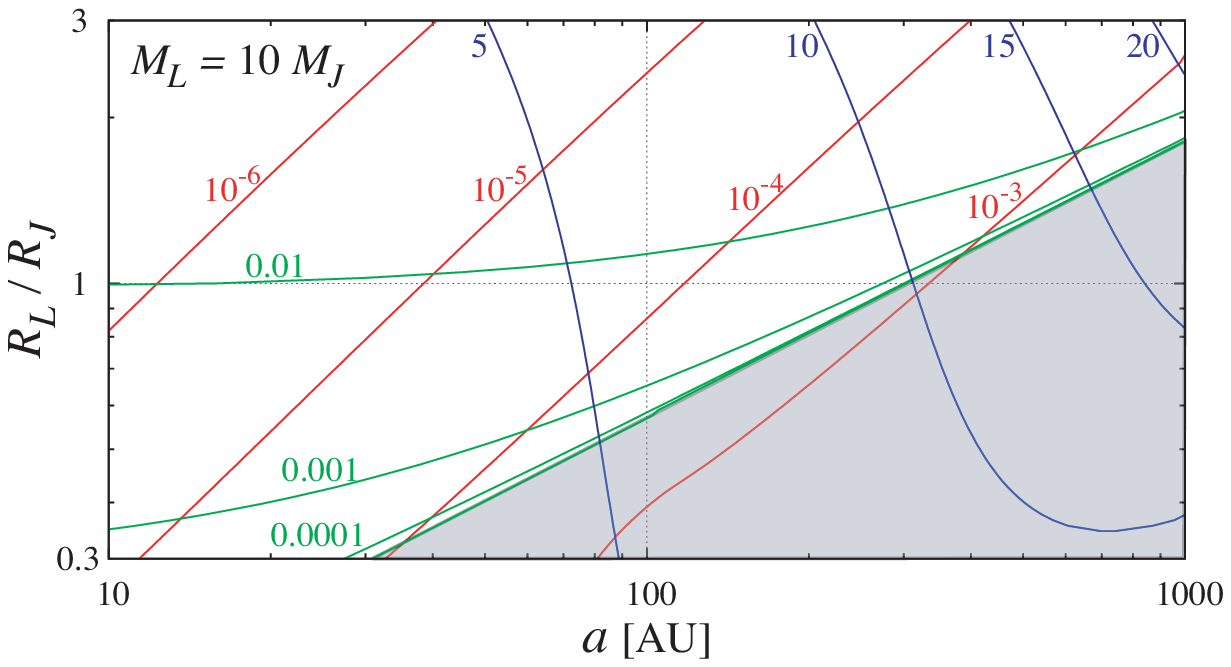} \\ %{m10.eps} 
\includegraphics[width=0.9\columnwidth]{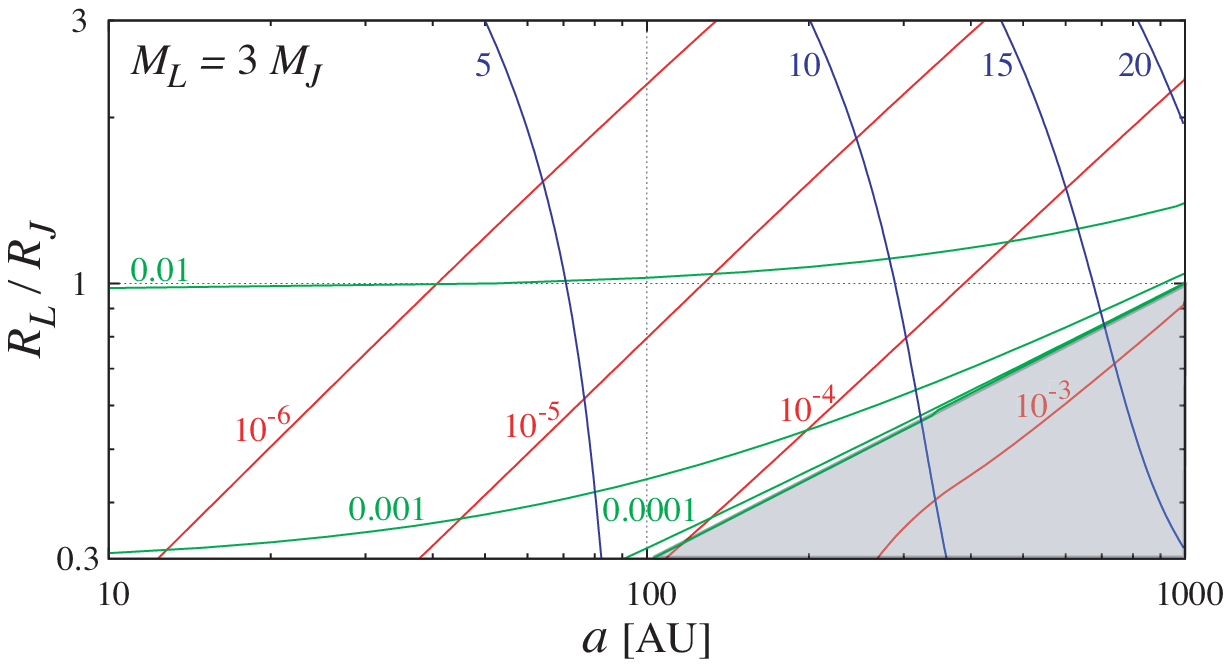} \\ %{m03.eps} 
\includegraphics[width=0.9\columnwidth]{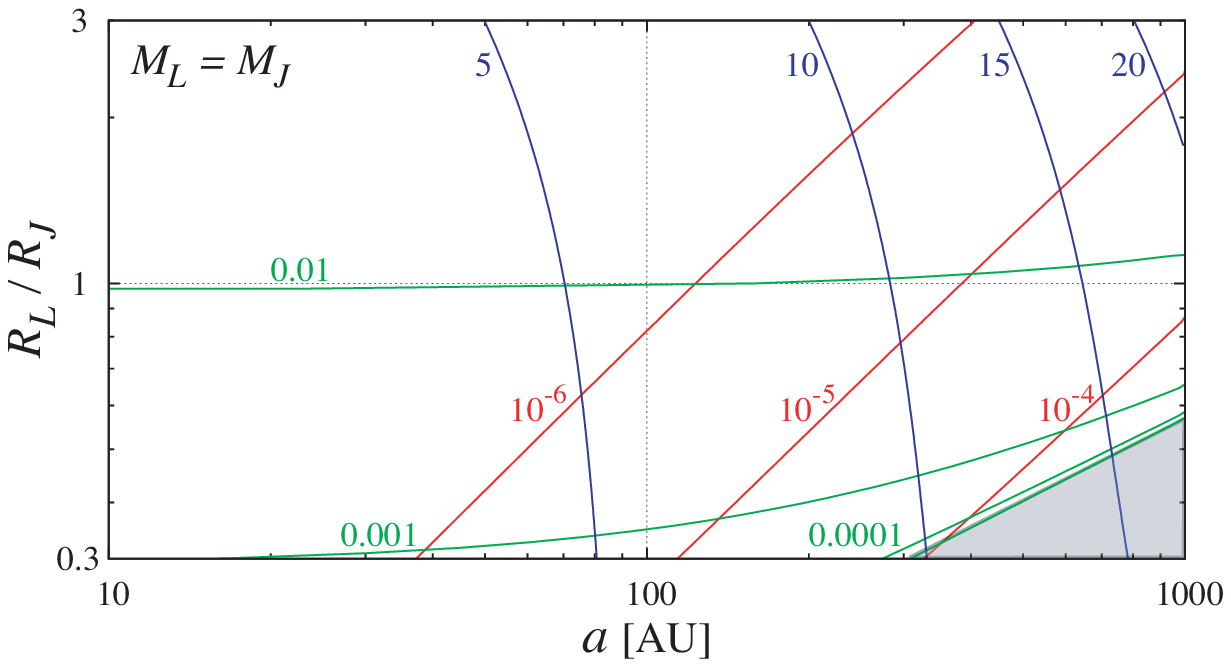}    %{m01.eps}
\end{tabular}
\caption{Sections of the isosurfaces of the rising spike $\xi$ (red), the depth $\Delta$ (green), 
and the duration $\tau$ in days (blue) for $M_L/M_J=10$, 3, and 1, from the top to
the bottle panels, respectively. Shaded region represents where the hollow disappears
$\Delta<0$, which we do not consider.}
\label{fig:iso_m}
\end{figure}

Figure~\ref{fig:iso_m} shows sections of the isosurfaces of $\Delta$, $\tau$, and $\xi$ for
$M_L=10M_J$, $3M_J$, and $M_J$ from the top panel to the bottom, respectively.
Red, green, and blue lines represent the sections of isosurfaces of the rising spike $\xi$, 
the depth $\Delta$, and the duration $\tau$, respectively. Shaded region denotes where 
$\Delta < 0$, which we do not consider because we focus on the transit observation. 
Since their parameter dependences look quite different in the section with constant $M_L$,
one could derive the physical parameters $M_L$, $R_L$, and $a$ from the figure.
For other sections, see App. B.

%%%%%%%%%%%%%%%%%%%%%%%%%%%%%%%
%%%%%%%%%%%%%%%%%%%%%%%%%%%%%%%

As can be seen in Fig.~\ref{fig:iso_m}, the rising spike is as large as $\sim 10^{-4}$, which  
can be measured in high precision observation such as the {\it Kepler} \citep{Jenkins2010},
for Jupitar-like planets with the orbital radius of a few hundred au . 
On the other hand, $\xi = 7\times 10^{-9} - 7\times 10^{-7}$ for $a=100 -1000$ au for
those super-earth planets with $M\sim10M_\oplus$ and $R\sim 3 R_\oplus$. Therefore,
it has no hope to see the rising spike in these cases.

In order to observe the rising spike, it must last long enough in addition to the height
larger than $10^{-4}$. Actually, the duration of the transit is about 8 days transit, 
while the rising spike becomes half the maximum height in 4 hours for  $M_L=10M_J$, 
$R_L= R_J$, and $a=200$ au (See Figs.~\ref{fig:LC} or \ref{fig:LD}). Therefore,
it could well be observed by {\it Kepler} with the long cadence mode, which has
the time resolution of 30 minutes.

%%%%%%%%%%%%%%%%%%%%%%%%%%%%%%%%%%%%%%%%%
\section{Limb darkening}
Realistically, the surface brightness of the source star is not uniform, but has position
dependence such that the centre is brighter than the edge. This phenomenon is called
the limb darkening of the star. Therefore, one must take this effect into account when 
any physical consequence is drawn from the light curve of the transit 
\citep{Agol2002, Agol2003}. In particular, one may wonder if the rising spike would be 
round or totally disappear. In this section, we show that this is not the case.

\begin{figure}
\includegraphics[width=0.9\columnwidth]{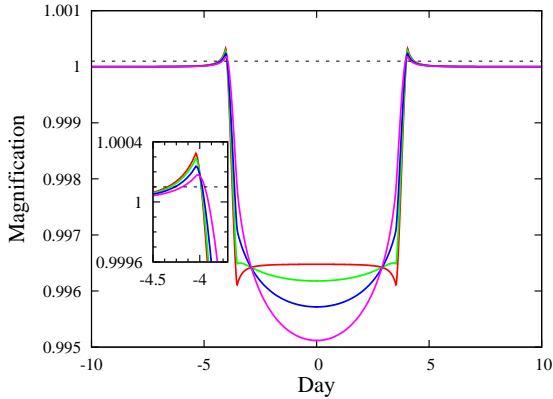}%{LD.eps}
\caption{Limb darkened light curves. We take the limb darkening parameters as 
($a_1$, $ a_2$)$=$ (0, 0), (0.18, 0.10), (0.40, 0.25), and (0.65, 0.35), respectively from the top 
to the bottom at the centre of the transit. Other parameters are set the same as in Fig.~\ref{fig:LC}.
In-figure is an enlargement of the rising spikes. Dashed line denotes ${\cal A}=1+10^{-4}$.}
\label{fig:LD}
\end{figure}

The effect of the limb darkening can be parametrize quadratically as
\begin{equation}
f_{LD}(\cos\mu) = 1- a_1 (1-\cos\mu) - a_2 (1-\cos\mu)^2,
\label{LD}
\end{equation}
where $\mu$ is an angle between the normal of the star surface and the line of the sight, and
$a_1$ and $a_2$ are constants of $O(0.1)$ \citep{Claret1995}. 
We can calculate the light curve of the transit 
including limb darkening effects from Eq.(\ref{mag}) with the limb darkening function (\ref{LD})
being inserted into the integrand in both the numerator and the denominator. We show
the transit light curve with the typical limb darkening in Fig.~\ref{fig:LD}. Here we plot the case
with ($a_1$, $a_2$)= (0.18, 0.10), (0.40, 0.25), and (0.65, 0.35) for $M_L= 10 M_J$, 
$R_L=10R_J$, and $a=200$ au. One can see that the amplitude of the rising spike would
diminish but still be large enough to be observed with the precision of $10^{-4}$ achieved 
by {\it Kepler} \citep{Jenkins2010} even for a large effect of limb darkening.

%%%%%%%%%%%%%%%%%%%%%%%%%%%%%%%
%%%%%%%%%%%%%%%%%%%%%%%%%%%%%%%

A comment is in order. Gravitational lensing effects are also seen as a bump at the 
bottom of the light curve if the limb darkening is negligible. Although it is another peculiar
feature of the lensing, the bump will totally disappear when the limb darkening effect is taken
into account, contrary to the rising spike (Agol2003). We thus consider the rising spike as 
the unique apparent feature of the gravitational lensing effect on the light curve.

%%%%%%%%%%%%%%%%%%%%%%%%%%%%%%%%%%%%%%%%%%
\section{Conclusion}
We have studied the gravitational lensing effects on the light curve of the transit which an 
extrasolar planet plays the role of the lens. The lensing effect becomes sizable only when
the orbital radius of the planet becomes larger than $\sim 10$ au. In such situations, the orbital 
velocity and/or orbital period are hard to be measured, and it thus seems more difficult to
obtain physical informations of the planet. Therefore, it is very important to
extract them from the gravitational lensing effects. We have claimed that the rising spike, the
sharp rise just before the ingress (or after the egress) of the transit, could be an excellent probe
for finding planet parameters. It has a peculiar feature that the height has different dependence
on the mass $M_L$, the radius $R_L$, and the orbital radius $a$ from other observables,
such as the depth and the duration of the transit. 

The rising spike can be large enough to be measured in high precision observation such as
the {\it Kepler}, typically when the orbital radius is a few hundred au. 
It can be observed even if we take into account the limb darkening effect in spite of the fact 
that other peculiar features of the lensing effect on the transit light curve will be lost.

Since the orbital radius is relatively large, one has at most one chance to observe in one's 
lifetime. The observability of this kind of the transit could be estimated optimistically as 
\begin{eqnarray}
& & \hspace{-7mm}
P  \sim \frac{t_{\rm obs}}{T_{\rm orbit}}\times \frac{R_S}{a} \times N_S 
\nonumber \\
& & \hspace{-3.5mm}
\sim \frac{\rm 4 \ yr}{3\times 10^3 \ {\rm yr}} 
\times \frac{{\rm R}_\odot}{\rm 200 \ au} \times 10^5
\sim 3\times 10^{-3},
\end{eqnarray}
where the first factor is the chance to meet the transit, the second is the probability of
the edge-on to the line of the sight, and the last is the number of the stars which will
be searched in the {\it Kepler} in four years of its operation. Although it seems unlikely to
find one by {\it Kepler}, one may observe in the future by using deeper and larger survey.

%%%%%%%%%%%%%%%%%%%%%%%%%%%%%%%%%%%%%%%%%%%%
\section*{Acknowledgments}
The work of M.H. is supported by the Grant-in-Aid for Scientific Research from the
Ministry of Education, Science, Sports, and Culture of Japan, No.~21740141.

%%%%%%%%%%%%%%%%%%%%%%%%%%%%%%%%%%%%%%%%%%%%%

%%%%%%%%%%%%%%%%%%%%%%%%%%%%%%%%%%%%%%%%%%%%%%

%%%%%%%%%%%%%%%%%%%%%%%%%%%%%%%%%%%%%%%
\appendix

%%%%%%%%%%%%%%%%%%%%%%%%%%%%%%%%%%%
\section{Three cases of the microlensing and the occultation}
\begin{figure}
\includegraphics[width=0.9\columnwidth]{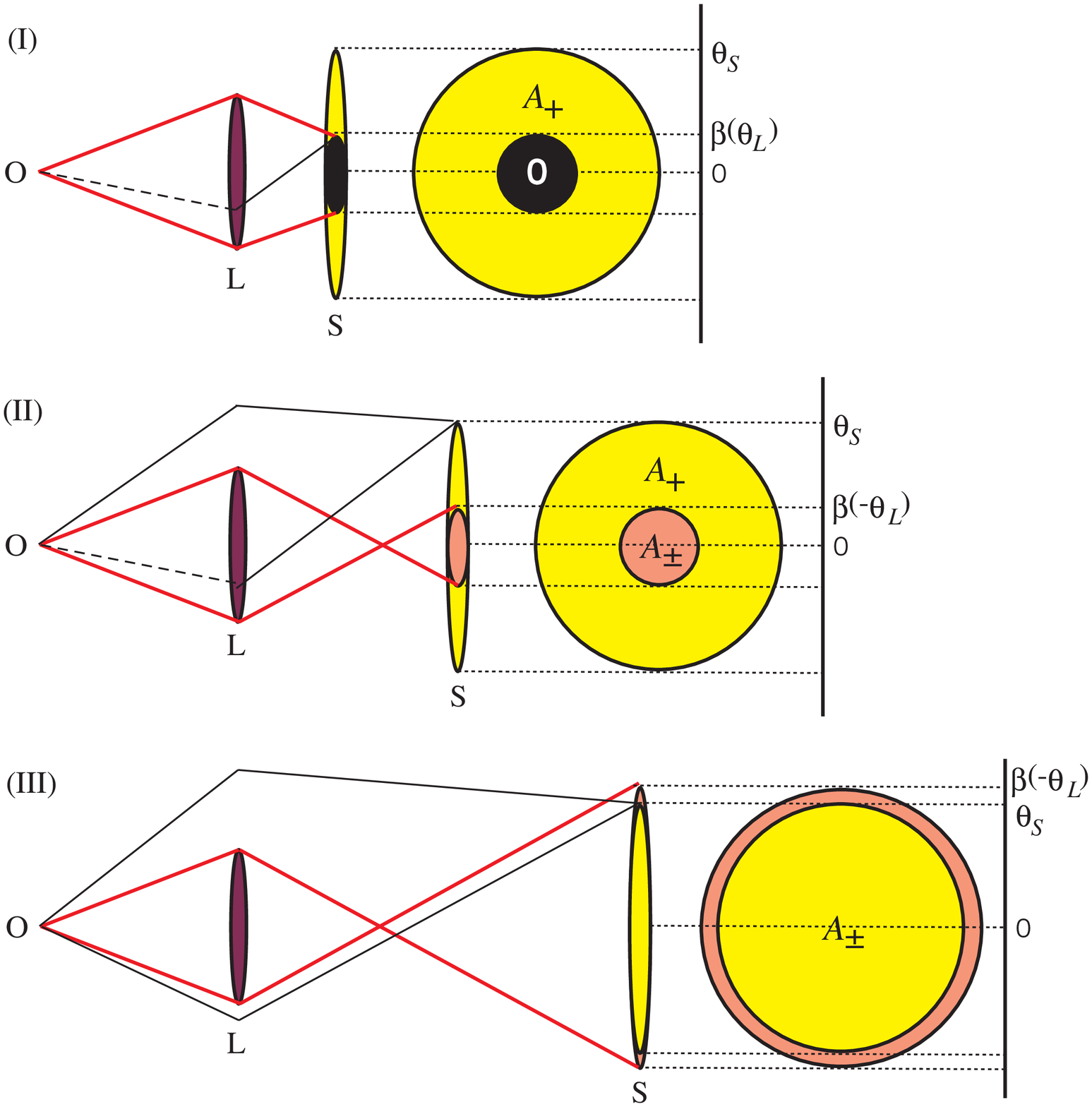}%{three.eps}
\caption{Schematic view of the microlensing and the occultation for three cases: 
Full, partial, and no occultations from the top to the bottom panels, respectively.}
\label{fig:three}
\end{figure}

Here we complete the cases of the microlensing and the occultation by the lens planet when
the centres of the planet and the star are aligned. The total magnification is given by Eq.(\ref{mag}),
and its denominator is obtained as Eq.(\ref{denominator}). There are three cases for the microlensing
and the occultation: (I) Full, (II) partial, and (III) no occultations, as shown in Fig.~\ref{fig:three}.
Full occultation occurs when $\theta_L > \theta_E$, where both $+$ and $-$ paths are obstructed,
while only $-$ path is shielded in the outer region. Therefore, the numerator of Eq.(\ref{mag}) can
be written as
\begin{eqnarray}
& & \hspace{-7mm}
\int_0^{2\pi} \int_{\beta(\theta_L)}^{\theta_S} A_+(\beta) \beta d\beta d\phi \nonumber \\
& & = \frac{\pi}{2}\left[\theta_S^2 - \beta_L^2 + \theta_S\sqrt{\theta_S^2+4\theta_E^2}
-\beta_L\sqrt{\beta_L^2+4\theta_E^2}\right],
\end{eqnarray}
where $\beta_L\equiv \beta(\theta_L)$. In the case (II), only partial occultation takes place 
in the outer region, where $\beta(-\theta_L) < \beta < \theta_S$, so that the numerator 
of Eq.(\ref{mag}) is given by
\begin{eqnarray}
& & \hspace{-7mm}
\int_0^{2\pi} \int_0^{\beta(-\theta_L)} A(\beta) \beta d\beta d\phi
+ \int_0^{2\pi} \int_{\beta(-\theta_L)}^{\theta_S} A_+(\beta) \beta d\beta d\phi \nonumber \\
& & = \frac{\pi}{2}\left[\theta_S^2 - \beta_L^2 + \theta_S\sqrt{\theta_S^2+4\theta_E^2}
-\beta_L\sqrt{\beta_L^2+4\theta_E^2}\right].
\end{eqnarray}
In the last case, there is no occultation, which is just the same situation as the point mass lens.
Thus, the numerator of Eq.(\ref{mag}) becomes
\begin{equation}
\int_0^{2\pi} \int_0^{\theta_S} A(\beta) \beta d\beta d\phi 
= \pi \theta_S\sqrt{\theta_S^2+4\theta_E^2}.
\end{equation}
Therefore, the magnification is obtained as
\begin{eqnarray}
& & \hspace{-7mm}
{\cal A}_{I,II} = \frac{1}{2}\left[ 1- \tilde{\beta}_L^2 + \sqrt{1+4\tilde{\theta}_E^2}
-\tilde{\beta}_L \sqrt{\tilde{\beta}_L^2+4\tilde{\theta}_E^2}\right], \\
& & \hspace{-7mm}
{\cal A}_{III} = \sqrt{1+4\tilde{\theta}_E^2},
\end{eqnarray}
where the tilde denotes the variable normalised with respect to $\theta_S$ such that
$\tilde{\beta}_L=\beta_L/\theta_S$ and $\tilde{\theta}_E=\theta_E/\theta_S$.

%%%%%%%%%%%%%%%%%%%%%%%%%%%%%%%
%%%%%%%%%%%%%%%%%%%%%%%%%%%%%%%

%%%%%%%%%%%%%%%%%%%%%%%%%%%%%%%%%%%%
\section[]{Other sections of the isosurfaces of the rising spike, the depth, and the duration}
\begin{figure}
\begin{tabular}{c}
\includegraphics[width=0.9\columnwidth]{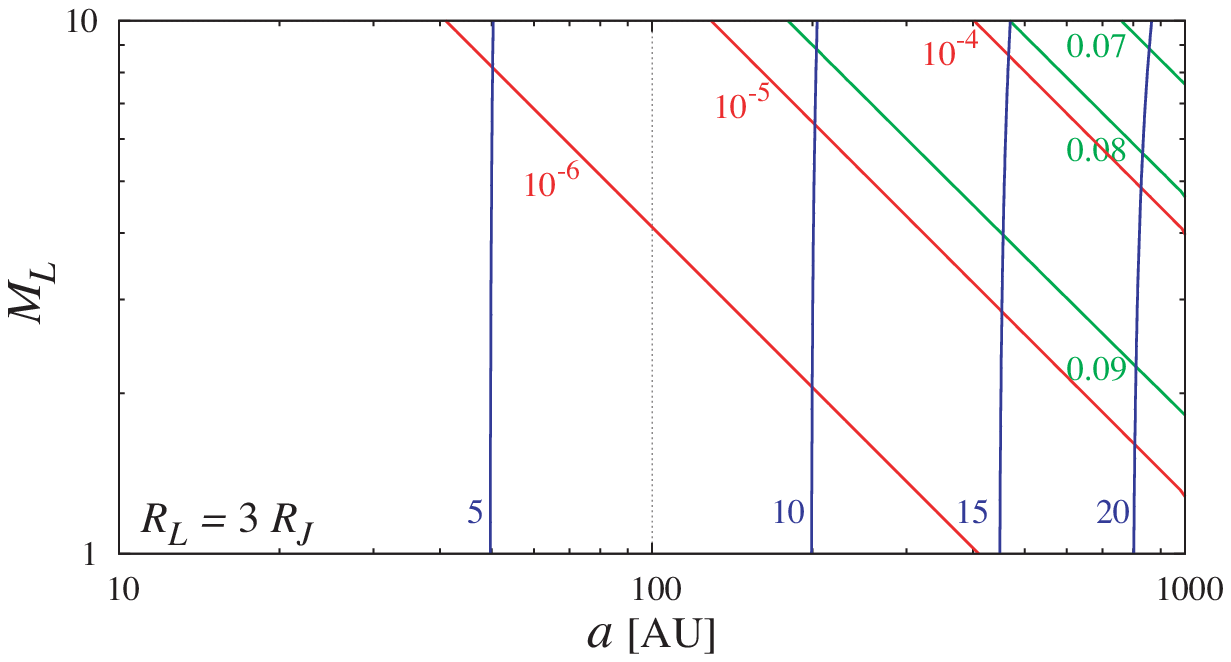} \\ %{r30.eps} 
\includegraphics[width=0.9\columnwidth]{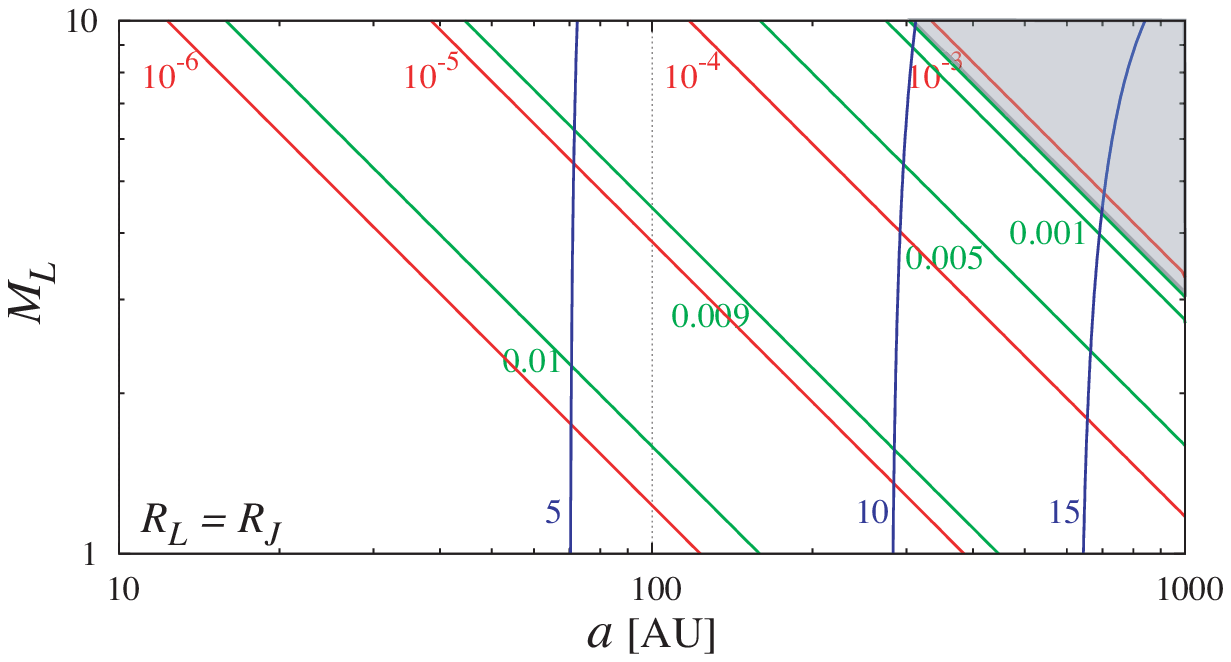} \\%{r10.eps} 
\includegraphics[width=0.9\columnwidth]{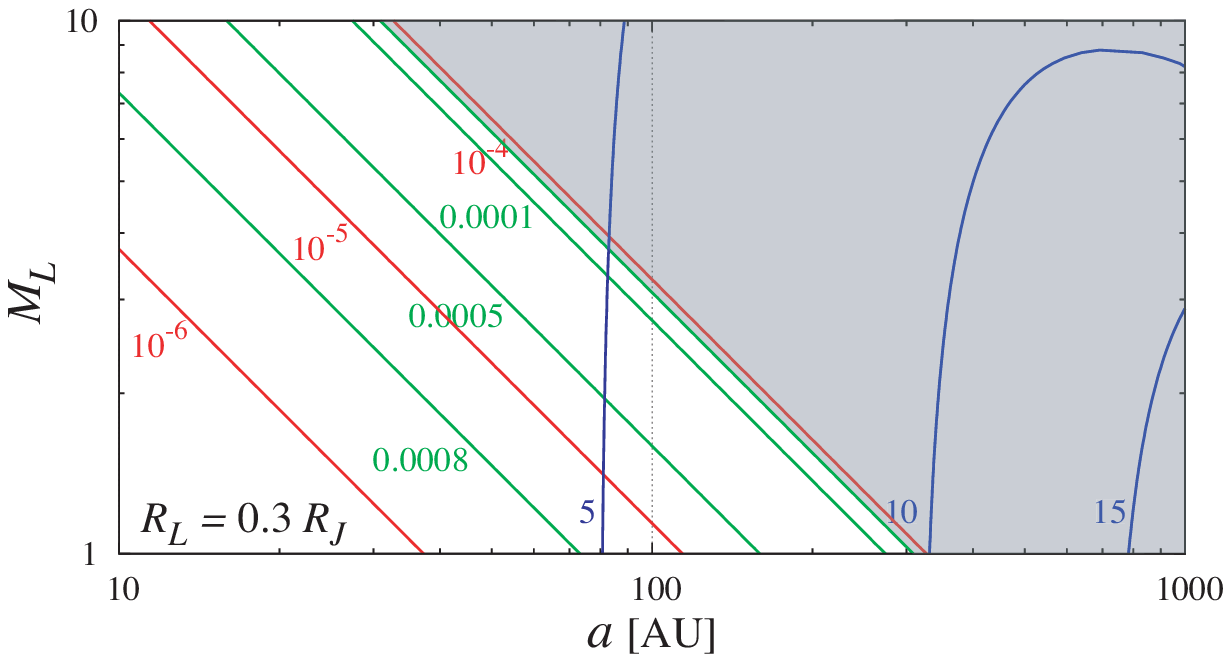}   %{r03.eps}
\end{tabular}
\caption{Sections of the isosurfaces of the rising spike $\xi$ (red), the depth $\Delta$ (green), 
and the duration $\tau$ in days (blue) for $R_L/R_J=3$, 1, and 0.3, from the top to
the bottle panels, respectively. Shaded region represents where the hollow disappears
$\Delta<0$, which we do not consider.}
\label{fig:iso_r}
\end{figure}

\begin{figure}
\begin{tabular}{cc}
\includegraphics[width=0.5\columnwidth]{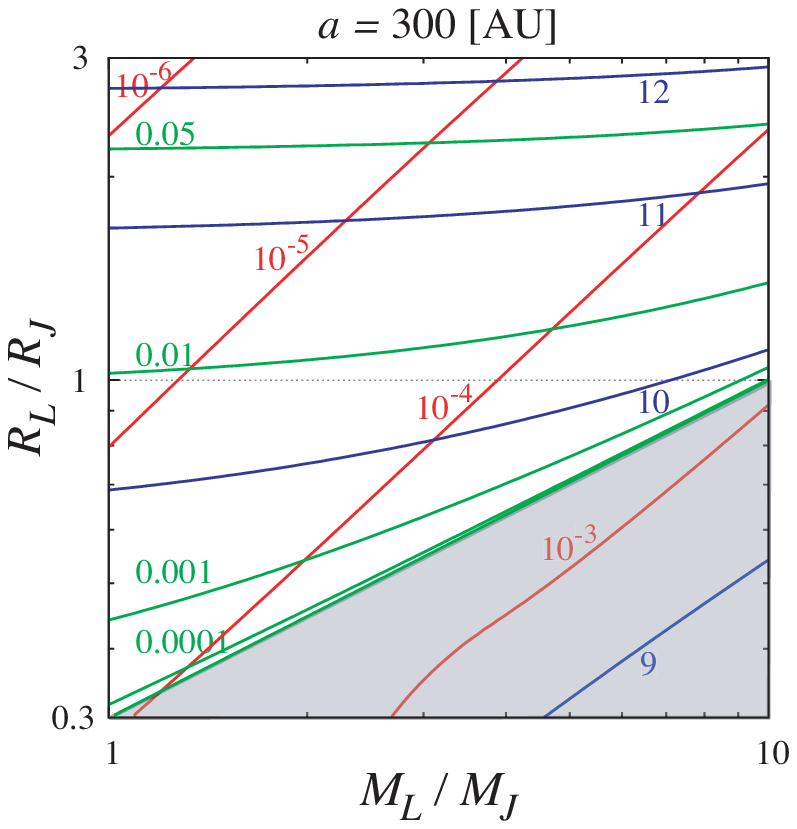} & %{a300.eps} 
\includegraphics[width=0.5\columnwidth]{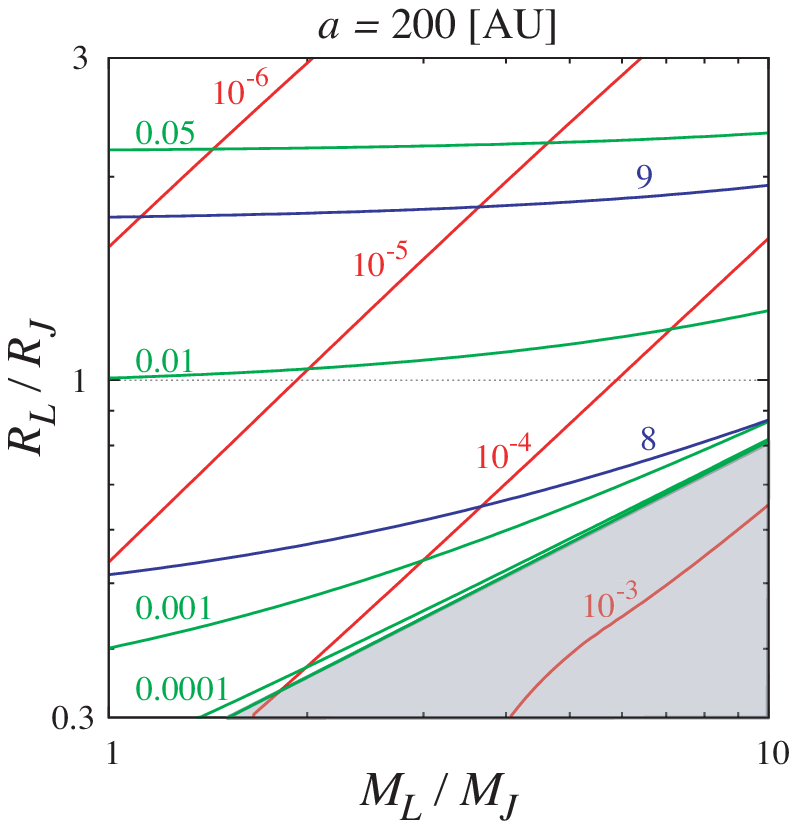} \\ %{a200.eps} 
\includegraphics[width=0.5\columnwidth]{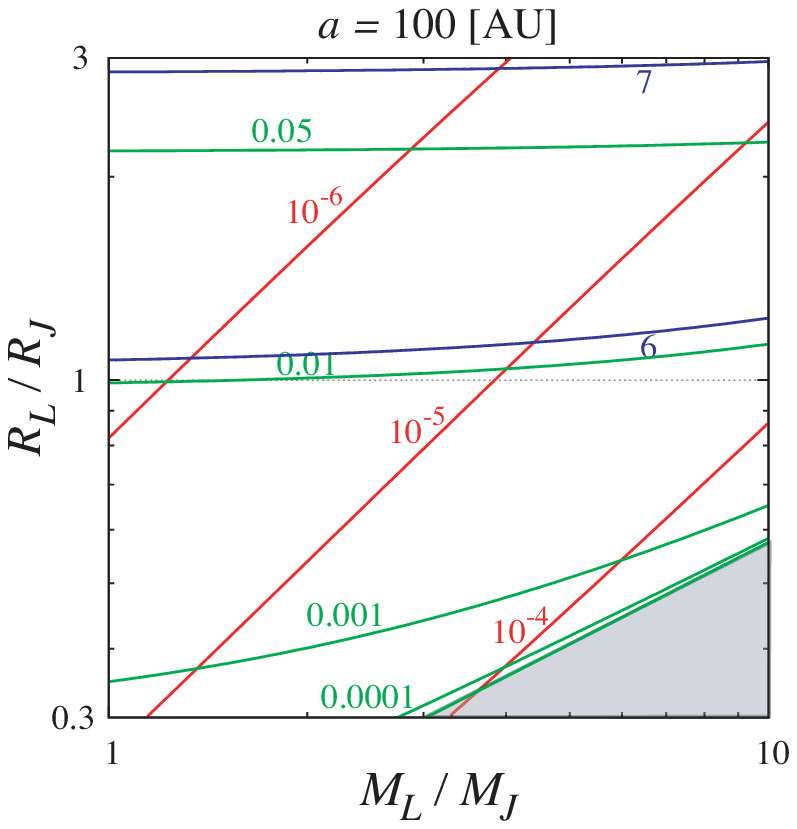} & %{a100.eps}
\includegraphics[width=0.5\columnwidth]{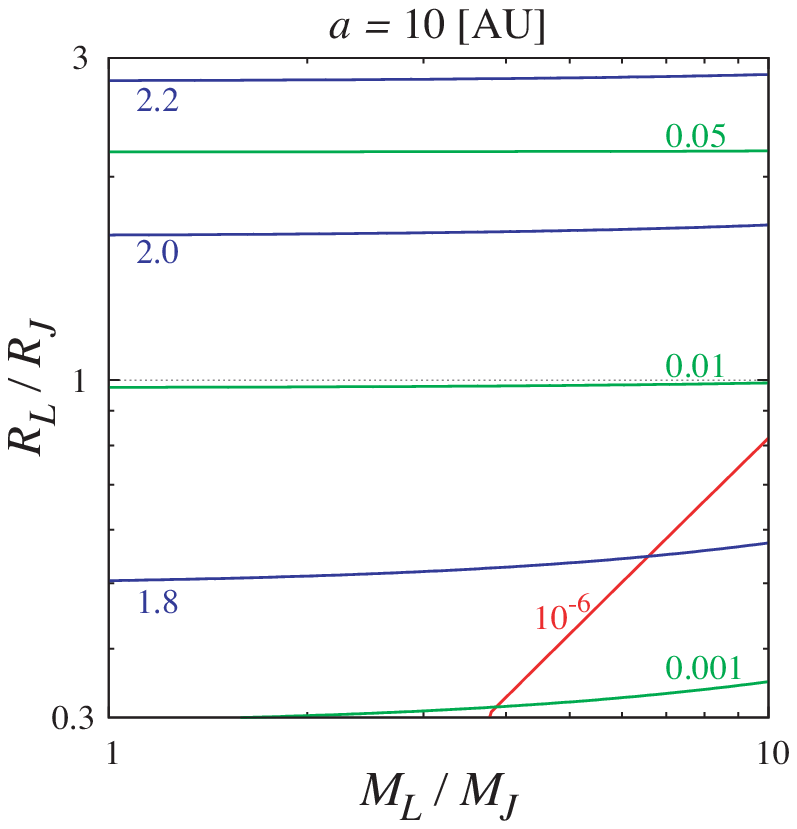}    %{a10.eps}
\end{tabular}
\caption{Sections of the isosurfaces of the rising spike $\xi$ (red), the depth $\Delta$ (green), 
and the duration $\tau$ in days (blue) for $a=300$, 200, 100, and 10 au. Shaded region 
represents where the hollow disappears $\Delta<0$, which we do not consider.}
\label{fig:iso_a}
\end{figure}

The rising spike $\xi$, the depth $\Delta$, and the duration $\tau$ of the transit are functions
of three variables: the mass $M_L$, the radius $R_L$, and the orbital radius $a$ of the planet.
It might be better to draw these isosurfaces in three dimensions, but may be difficult to look at. 
Therefore, we show some constant $M_L$ sections in Fig.~\ref{fig:iso_m}, because they seem
to have very different curves in those sections. Here we show constant $R_L$ and $a$ sections
for completeness in Figs.~\ref{fig:iso_r} and \ref{fig:iso_a}, respectively. In these figures, two of 
the observables ($\xi$ and $\Delta$ in Fig.~\ref{fig:iso_r}, while $\Delta$ and $\tau$ in 
Fig.~\ref{fig:iso_a}) have similar dependence on the physical parameters. Notice that one of 
these parameters ($\Delta$ in Fig.~\ref{fig:iso_r}, while $\tau$ in Fig.~\ref{fig:iso_a}) changes 
very rapidly with respect to the parameter perpendicular to the two shown in the figures.

%%%%%%%%%%%%%%%%%%%%%%%%%%%%%%%
%%%%%%%%%%%%%%%%%%%%%%%%%%%%%%%

%%%%%%%%%%%%%%%%%%%%%%%%%%%%%%%
%%%%%%%%%%%%%%%%%%%%%%%%%%%%%%%

\label{lastpage}
\end{document}